\documentclass[aps,prd,superscriptaddress,nofootinbib]{revtex4-2}

\usepackage[T1]{fontenc}
\usepackage[utf8]{inputenc}
\usepackage{lmodern}
\usepackage{amsmath}
\usepackage{amssymb}
\usepackage{mathrsfs}
\usepackage{booktabs}
\usepackage[compat=1.1.0]{tikz-feynman}
\usepackage{graphicx}
\usepackage{bm}
\usepackage{hyperref}
\hypersetup{colorlinks=true, citecolor=blue, urlcolor=blue, linkcolor=blue}

\begin{document}

\title{A mechanism of the power-counting breaking in chiral perturbation theory}

\author{Ming-Ze Ma}
\author{Shao-Zhou Jiang}
\email{jsz@gxu.edu.cn}
\affiliation{Key Laboratory for Relativistic Astrophysics, School of Physical Science and Technology, Guangxi University, Nanning 530004, People's Republic of China}

\begin{abstract}
Chiral perturbation theory is a power tool for studying the low-energy strong interactions. In the conventional chiral expansion, according to the power counting rule, the higher-order corrections for physical observables are expected to decrease order by order as the chiral dimension increases. However, most previous study of $\pi K$ scattering lengths by different references showed that the $\mathcal{O}(p^6)$ correction to the scattering lengths is anomalously large and even exceeds the $\mathcal{O}(p^4)$ one. It poses a serious challenge to the convergence of chiral expansion. In order to explore the underlying origin of this phenomenon, the contributions of two-loop diagrams, one-loop diagrams and tree diagrams for the $\pi K$ scattering amplitude are detailed analyzed. The reducible two-loop integrals, sunset integrals and vertex integrals from the $\mathcal{O}(p^6)$ order all violate the power counting rule. The sunset and vertex integrals can be largely canceled numerically, but not enough. Furthermore, $\pi K$ scattering is compared with $\pi\pi$ scattering, which seems not violate the power counting rule. The analysis indicates that $\pi\pi$ scattering contains a higher symmetry, although each $\mathcal{O}(p^6)$-order part also violates the power counting rule, but the violation is much smaller than the one from $\pi K$ scattering. Large kaon mass also makes a contribution to the power-counting breaking, but it is not a major origin. Hence, it needs to pay more attention to the many-loop low-symmetry diagrams in future work. These diagrams may give large contribution and cannot be ignored.
\end{abstract}

\maketitle

\section{Introduction}

Quantum chromodynamics (QCD) is the fundamental theory of the strong interaction. In the low-energy region
$(\lesssim 1\ \mathrm{GeV})$, QCD becomes strongly non-perturbative, and the conventional perturbative methods based on the quark and gluon degrees of freedom cease to be applicable. In order to describe strong-interaction dynamics effectively in this region, chiral perturbation theory (ChPT), which is based on the chiral symmetry $SU(3)_L \times SU(3)_R$ spontaneous breaking to $SU(3)_V$, is developed and has gradually become the common effective field theory framework for low-energy QCD phenomena \cite{weinberg1979, Gasser:1983yg, Gasser:1984gg}. In this framework, the pseudo-Goldstone bosons generated by spontaneous symmetry breaking are treated as the mesonic degrees of freedom, i.e. $\pi$, $K$, and $\eta$. Actually, the mass of both the light quark $m_q$ and pseudoscalar mesons $m_M$ are small compared with the chiral scale $\Lambda_\chi\approx 4\pi F_\pi \approx 1\ \mathrm{GeV}$. It is reasonable to regard $m_q$ and $m_M$ as zeros to some extent. Hence, the chiral symmetry $SU(3)_L \times SU(3)_R$ is approximately correct. When the external momentum $p$ is also small, the effective Lagrangian and the scattering amplitudes can be organized systematically according to chiral order $n$, which is conventionally denoted by $\mathcal{O}(p^n)$. The hierarchy of the chiral series is controlled by the dimensionless expansion parameters $p/\Lambda_\chi$, i.e., one chiral dimension makes a contribution that is about $p/\Lambda_\chi$ times as much. The parameter $p/\Lambda_\chi$ provides a rough estimate of the calculation accuracy and assumes the convergence of the chiral series. The detailed procedure is the Weinberg power counting rule \cite{weinberg1979}. It estimates the contribution of each diagram. In short, for an arbitrary Feynman diagram, the chiral dimension $D$ satisfies
\begin{equation}
D = 2 + 2L + \sum_{i} (d_i - 2), \label{eq:pc}
\end{equation}
where $L$ denotes the number of loops and $d_i$ denotes the chiral dimension of each vertex.
More detailed introduction to ChPT and the power counting rule can be found in Refs. \cite{pich1998effectivefieldtheory,Scherer:2002tk,scherer2005chiralperturbationtheoryprimer,Ecker:1994gg,Leutwyler_1994,Scherer_2010}.

However, the power counting rule only estimates one Feynman diagram at a time. It might not be very precise. During the actual application of ChPT, one should pay more attention to all Feynman diagrams at a given chiral dimension. Ref.~\cite{Bijnens:2014lea} gives a typical ratio at each order, $\mathcal{O}(p^4)\sim 25\%$, $\mathcal{O}(p^6)\sim 7\%$ and $\mathcal{O}(p^8)\sim 1.5\%$. If these ratios are approximately satisfied, the ChPT calculation is quite reliable. One can easily estimate the theoretical error, and thereby determine at what level the calculation should be carried out for a specific problem. For example, if a precision of 10\% is already sufficient, the $\mathcal{O}(p^6)$ calculation is not needed. In addition, the unnecessary high-order and high-precision calculation is usually tedious and boring, many unknown high-order low-energy constants (LECs) will also appear, which will reduce the predictive power of ChPT. A reliable estimate can avoid this unnecessary calculation. If the power counting rule works well, the above ratios can roughly serve as the basis for the calculation. If not, a more detailed analysis is required. The theoretical error should be appropriately increased, or more complex higher-order calculations should be introduced. Hence, this paper mainly pays attention to the power counting of the sum of all diagrams at a given order, but not a special Feynman diagram.

For most processes, the power counting rule works well.  One of the most representative successful examples is the $\pi\pi$ scattering lengths $a_0^0$ and $a_0^2$~\cite{Colangelo_2001,Bijnens_1996,Bijnens_1997,Colangelo2001piPS,Leutwyler:2006qq,Colangelo_2000,Bijnens:2004eu}. Table~\ref{tab:scattering_lengths} lists the results by different global-fit methods~\cite{Bijnens:2014lea,Yang:2020eif,Pan:2023qja}. The values of $a_0^0$ and $a_0^2$ show that the $\mathcal{O}(p^2)$ contributions dominate the most contribution, while the $\mathcal{O}(p^4)$ and $\mathcal{O}(p^6)$ corrections remain successively smaller. This pattern is fully consistent with the estimated ratios in Ref.~\cite{Bijnens:2014lea}.
The other examples, such as mass ratio $m_s/\hat m$~\cite{Amoros:1999dp,Amoros:2000mc,Bijnens:2011tb}, $K_{\ell4}$ form factors~\cite{Amoros:2000mc}, pion scalar radius~\cite{Bijnens:2003xg} and so on, also satisfy the power counting rule~\cite{Bijnens:2014lea,Yang:2020eif,Pan:2023qja}, too.

\begin{table}[htbp]
\centering
\caption{The $\pi\pi$ ($a_0^0$, $a_0^2$) and $\pi K$ ($a_0^{1/2}$, $a_0^{3/2}$) scattering lengths at each order}
\label{tab:scattering_lengths}
\begin{tabular}{cccc|ccc|ccc}
\hline\hline
Scattering & $\mathcal{O}(p^2)$ & $\mathcal{O}(p^4)$ & $\mathcal{O}(p^6)$ & $\mathcal{O}(p^2)$ & $\mathcal{O}(p^4)$ & $\mathcal{O}(p^6)$ & $\mathcal{O}(p^2)$ & $\mathcal{O}(p^4)$ & $\mathcal{O}(p^6)$ \\
lengths&\multicolumn{3}{c|}{Ref.~\cite{Bijnens:2014lea}}&\multicolumn{3}{c|}{Ref.~\cite{Yang:2020eif}}&\multicolumn{3}{c}{Ref.~\cite{Pan:2023qja}} \\
\hline
$a_0^0$ & 0.160 & 0.044 & 0.012 & 0.1592 & 0.0453 & 0.0098 & 0.159 & 0.045 & 0.015 \\
$10a_0^2$ & -0.456 & 0.016 & -0.001 & -0.455 & 0.022 & -0.010 & -0.455 & 0.020 & 0.003 \\
$a_0^{1/2}M_\pi$ & 0.142 & 0.031 & 0.051 & 0.142 & 0.033 & 0.049  & 0.142 & 0.034 & 0.049 \\
$10a_0^{3/2}M_\pi$ & -0.710 & 0.070 & 0.160 & -0.709 & 0.094 & 0.142 & -0.709 & 0.093 & 0.179 \\
\hline\hline
\end{tabular}
\end{table}

However, some observables give a bad convergence, the power counting rule seems to be broken. For example, Table~\ref{tab:scattering_lengths} also gives the $\pi K$ scattering lengths $a_0^{1/2}$ and $a_0^{3/2}$, but they have a bad convergence. Their $\mathcal{O}(p^6)$ contributions are larger than the $\mathcal{O}(p^4)$ ones, although $\pi K$ scattering seems very similar to $\pi\pi$ scattering. This example indicates that the power counting rule indeed is just an estimate, but not a universal rule.
Besides these $\pi K$ scattering lengths, the other processes also exist power-counting breaking. The two-loop calculation of $\eta \rightarrow 3\pi$ shows that the $\mathcal{O}(p^6)$ correction remains large \cite{Bijnens_2007}. The expected order-by-order suppression is therefore weakened. Further studies have questioned the applicability of the standard chiral expansion for $\eta \rightarrow 3\pi$ and have argued that resummation may be required \cite{Kolesar:2016jwe}.  All the above example indicates that the diagrams classified according to Eq. \eqref{eq:pc} might have some problem. It requires further research. For the baryon sector, the power-counting breaking is often even more pronounced \cite{Becher:1999he,Fuchs:2003qc}. Of course, this is another story and is not mentioned in this paper.

Hence, a systematic analysis of the mathematical structure and physical origin of the chiral power-counting breaking is therefore important. One of the major problems of low-energy QCD is error estimation. Although different approximations can give a theoretical prediction, most of them cannot give a reliable error. ChPT can inherently estimate the truncation term by the power counting rule. If the power counting rule is violated, the predictive power of ChPT will be greatly reduced. Hence, one would better know whether the error meets the requirements before making the calculation. If the higher-order contribution is large, the low-order calculation needs a large reliable error.

In order to address such difficulties, resummed chiral perturbation theory (RChPT) was proposed. Because most examples violating the power counting are related to the strange quark, RChPT attributes the breaking of the chiral expansion partly to the strong vacuum fluctuations induced by $s\bar{s}$ pairs. If the Zweig rule is strongly violated in the scalar channel, the higher-order terms that are usually treated as small corrections. It may become numerically important. RChPT does not impose the assumption that the $\mathcal{O}(p^2)$ contribution must be much larger than the $\mathcal{O}(p^4)$ and $\mathcal{O}(p^6)$ ones. Instead, RChPT reorganizes potentially unstable contributions, especially those related to the quark condensate and decay constants, by resummation \cite{Descotes-Genon:2007sqh,Descotes-Genon:2003qye}. At present, RChPT has achieved some success in the study of $\pi\eta$ scattering \cite{Koles_r_2008}. Of course, the resummed framework is also more complicated and involves more parameters. The recent lattice QCD results further show that vacuum fluctuations induced by strange quarks do exist, but the extreme scenario of a complete collapse of the $\mathcal{O}(p^2)$ contribution has not been observed \cite{FlavourLatticeAveragingGroupFLAG:2021npn}.

This paper only restricts in the conventional ChPT, because it has been applied widely. In fact, there might be many different mechanisms to violate the power counting rule. This paper is merely a preliminary exploration of the breaking mechanism. A comprehensive exploration is a huge project.
Against this background, $\pi K$ scattering provides a particularly suitable example for a detailed study of the power-counting breaking. A detailed analysis of $\pi K$ scattering can therefore clarify how the expected order-by-order hierarchy is lost in a concrete hadronic process. Hence, the present work takes $\pi K$ scattering lengths as an example and examines the possible origin of the power-counting breaking. In addition, $\pi\pi$ scattering lengths are also introduced, because it keeps the power counting rule and is similar to $\pi K$ scattering. It can be served as an excellent comparative example. The discussion is restricted in Ref.~\cite{Bijnens:2014lea}. Although the other tricks, such as another renormalization scheme, may mitigate the power-counting breaking, it is only wise after the event. Because before the high-order calculation, one cannot know which trick will be chosen.

This paper is organized as follows. Section \ref{sec:tf} briefly introduces the chiral Lagrangian and the calculation of the $\pi K/\pi\pi$ scattering amplitude and scattering lengths. In Sec. \ref{sec:num}, the numerical results have been separated into five parts, in order to reveal the possible aspects of the problem. Section \ref{sec:origin} gives a detailed analysis on the origin of the  power-counting breaking, form two-loop, one-loop to tree diagrams. This section also compares $\pi K$ scattering with $\pi\pi$ scattering, in order to find their difference. Section \ref{sec:con} concludes with a summary and gives a discussion.

\section{Theoretical Framework}\label{sec:tf}

\subsection{Chiral Lagrangian}

To organize the chiral expansion, the $SU(3)$ chiral Lagrangian is written as~\cite{Gasser:1983yg,Gasser:1984gg,Bijnens:1999sh,Bijnens:1999hw}
\begin{equation}
\begin{aligned}
\mathcal{L}_{\mathrm{eff}}
&=\mathcal{L}_2+\mathcal{L}_4+\mathcal{L}_6+\cdots \\
&=\frac{F_0^2}{4}\mathrm{Tr}\!\left(D_\mu U D^\mu U^\dagger\right)
+\frac{F_0^2}{4}\mathrm{Tr}\!\left(\chi U^\dagger+U\chi^\dagger\right)
+\sum_{i=1}^{10}L_i O_i^{(4)}
+\sum_{i=1}^{90} C_i O_i^{(6)}
+\cdots ,
\end{aligned}
\end{equation}
where $U(x)=\exp(i\phi/F_0)$ contains the pseudoscalar meson fields $\phi$, $F_0$ is the meson decay constant in the chiral limit. $\chi=2B_0M_q$, and $M_q=\mathrm{diag}(M_u,M_d,M_s)$, which is the quark masses matrix. The coefficients $L_i$ and $C_i$ denote the LECs at $\mathcal{O}(p^4)$ and $\mathcal{O}(p^6)$ orders, respectively. The $\mathcal{O}(p^4)$ and $\mathcal{O}(p^6)$ contact terms are omitted here, since they are not relevant to the discussion in this paper.

The mesonic chiral Lagrangian has been constructed systematically up to $\mathcal{O}(p^8)$ \cite{Bijnens:2018lez}. The present work, however, only focuses on the $\mathcal{O}(p^6)$ order, where the breaking of the power counting rule has already appeared explicitly in the $\pi K$ scattering.

\subsection{$\pi K$ scattering kinematics}

Following Ref.~\cite{Bijnens:2004bu}, the basic kinematics and threshold parameters of $\pi K$ scattering are introduced first, since they are related to the $\pi K$ scattering lengths discussed in this paper.

For the elastic scattering process
\begin{equation}
    \pi(p_1) + K(p_2) \rightarrow \pi(p_3) + K(p_4),
\end{equation}
the kinematics are fully characterized by the Mandelstam variables~\cite{Mandelstam:1958xc}
\begin{equation}
    s = (p_1 + p_2)^2, \qquad t = (p_1 - p_3)^2, \qquad u = (p_1 - p_4)^2,
\end{equation}
which satisfy
\begin{equation}
    s + t + u = 2m_\pi^2 + 2m_K^2.
\end{equation}

The pions have isospin $I=1$ and the kaons have isospin $I=1/2$. The full $\pi K$ scattering system can therefore be decomposed into two isospin eigenchannels, $I=1/2$ and $I=3/2$. In practical calculations, it is convenient to start from a specific process, such as $\pi^+K^+\rightarrow\pi^+K^+$. In particular, $\pi^+K^+\rightarrow\pi^+K^+$ corresponds to the pure $I=3/2$ channel, so it is much easier to handle. The full isospin structure can then be reconstructed by crossing relation. In this work, the discussion below is carried out with the $I=3/2$ channel as the explicit example. The result of the $I=1/2$ channel is related to the $I=3/2$ channel, as shown in Eq. \eqref{eq:t1232} below.

The total scattering amplitude can be expanded as
\begin{equation}
    T(s,t,u) = T_{2}(s,t,u) + T_{4}(s,t,u) + T_{6}(s,t,u) + \mathcal{O}(p^8).\label{eq:piKA}
\end{equation}
Here the subscripts denote the chiral dimensions of the corresponding contributions,
while $\mathcal{O}(p^8)$ denotes the higher-order terms not included in the present analysis. The two isospin amplitudes satisfy the crossing relation
\begin{equation}
    T^{\frac{1}{2}}(s,t,u)
    =-\frac{1}{2}T^{\frac{3}{2}}(s,t,u)
    +\frac{3}{2}T^{\frac{3}{2}}(u,t,s). \label{eq:t1232}
\end{equation}

To characterize the scattering kinematics, the magnitude of the center-of-mass three-momentum is introduced as
\begin{equation}
    q_{\pi K}^2
    =\frac{s}{4}
    \left(1-\frac{(m_K+m_\pi)^2}{s}\right)
    \left(1-\frac{(m_K-m_\pi)^2}{s}\right).
\end{equation}
The variables $t$ and $u$ can then be written as
\begin{equation}
    t=-2q_{\pi K}^2(1-\cos\theta),\qquad
    u=-s-t+2m_K^2+2m_\pi^2,
\end{equation}
where $\theta$ denotes the scattering angle of $\pi$ or $K$ in the center-of-mass frame. The amplitudes in the different isospin channels are expanded in partial waves as
\begin{equation}
    T^I(s,t,u)=16\pi\sum_{\ell=0}^\infty(2\ell+1)P_\ell(\cos\theta)t_\ell^I(s).
\end{equation}
Near the reaction threshold, the partial-wave amplitudes are written as
\begin{equation}
    t_\ell^I(s)
    =\frac{1}{2}\sqrt{s}\,q_{\pi K}^{2\ell}
    \left(a_\ell^I+b_\ell^I q_{\pi K}^2+\mathcal{O}(q_{\pi K}^4)\right),
    \label{eq:t}
\end{equation}
where $a_\ell^I$ and $b_\ell^I$ are the threshold parameters. In particular, $a_\ell^I$ denotes the scattering length. At the threshold,
\begin{equation}
    s=(m_\pi+m_K)^2,\qquad
    t=0,\qquad
    u=(m_\pi-m_K)^2.\label{eq:thpik}
\end{equation}
For the $S$ wave, $\ell=0$, the scattering lengths in the two isospin channels are
\begin{equation}
    a_0^I
    =
    \frac{
    T^I\bigl((m_\pi+m_K)^2,\,0,\,(m_\pi-m_K)^2\bigr)
    }
    {8\pi(m_\pi+m_K)},
    \qquad I=\frac12,\frac32.\label{eq:pikl}
\end{equation}

\subsection{$\pi\pi$ scattering kinematics}

Since the later discussion compares the behavior of $\pi K$ and $\pi\pi$ scattering lengths, this subsection also summarizes the corresponding framework of $\pi\pi$ scattering as well. The details can be found in Ref.~\cite{Bijnens:2004eu}.

Because the two external mesons are identical, the $\pi\pi$ scattering amplitude can be expressed in terms of a single function $A(s,t,u)$ that is symmetric under the interchange of $t$ and $u$,
\begin{equation}
A\bigl(\pi^a(p_1)\pi^b(p_2)\to\pi^c(p_3)\pi^d(p_4)\bigr)
=\delta^{ab}\delta^{cd}A(s,t,u)
+\delta^{ac}\delta^{bd}A(t,u,s)
+\delta^{ad}\delta^{bc}A(u,t,s).\label{eq:pipiA}
\end{equation}
The corresponding isospin amplitudes are
\begin{equation}
\begin{aligned}
T^0(s,t)&=3A(s,t,u)+A(t,u,s)+A(u,s,t),\\
T^1(s,t)&=A(t,u,s)-A(u,s,t),\\
T^2(s,t)&=A(t,u,s)+A(u,s,t),
\end{aligned}
\end{equation}
where the superscripts denote the isospin.

For the equal pion masses, the center-of-mass momentum is
\begin{equation}
q_{\pi\pi}^2=\frac{1}{4}\left(s-4m_\pi^2\right),
\end{equation}
and the scattering angular dependence of the Mandelstam variables becomes
\begin{equation}
t=-2q_{\pi\pi}^2(1-\cos\theta),\qquad
u=-2q_{\pi\pi}^2(1+\cos\theta).
\end{equation}

The standard partial-wave expansion of the $\pi\pi$ isospin amplitudes is written as
\begin{equation}
T^I(s,t)=32\pi\sum_{\ell=0}^{\infty}(2\ell+1)P_\ell(\cos\theta)\,t_\ell^I(s).
\end{equation}
Near threshold, the partial waves are expanded as
\begin{equation}
t_\ell^I(s)=q_{\pi\pi}^{2\ell}\left(a_\ell^I+b_\ell^I q_{\pi\pi}^2+\mathcal{O}(q_{\pi\pi}^4)\right),
\end{equation}
where $a_\ell^I$ and $b_\ell^I$ denote the threshold parameters. In particular, $a_0^0$ and $a_0^2$ are the $S$-wave scattering lengths in the $I=0$ and $I=2$ channels, respectively. They will be discussed later.

An important structural property of $\pi\pi$ scattering is that the inelasticity starts only at the $\mathcal{O}(p^8)$ order. Up to the $\mathcal{O}(p^6)$ order, only $t_0^0$, $t_1^1$, and $t_0^2$ are nonzero. As a consequence, the amplitude can be reconstructed up to the $\mathcal{O}(p^6)$ order in the form~\cite{Stern:1993rg}
\begin{equation}
\begin{aligned}
A(s,t,u)=&\,C(s,t,u)
+32\pi\Bigg(
\frac{1}{3}V^0(s)
+\frac{3}{2}(s-u)V^1(t)
+\frac{3}{2}(s-t)V^1(u) \\
&\qquad\qquad
+\frac{1}{2}V^2(t)
+\frac{1}{2}V^2(u)
-\frac{1}{3}V^2(s)
\Bigg),
\end{aligned}
\end{equation}
where the functions $V^I(s)$ contain the singularities associated with intermediate states of isospin $I$ in the different channels. They are satisfied
\begin{equation}
\mathrm{Im}\,V^{0,2}(s)=\mathrm{Im}\,t_0^{0,2}(s),\qquad
\mathrm{Im}\,V^1(s)=\frac{1}{s-4m_\pi^2}\,\mathrm{Im}\,t_1^1(s).
\end{equation}
Using $s+t+u=4m_\pi^2$, the polynomial part can be written as
\begin{equation}
C(s,t,u)=c_1+c_2 s+c_3 s^2+c_4 (t-u)^2+c_5 s^3+c_6 s(t-u)^2.
\end{equation}

For $\pi\pi$ scattering, the threshold point is
\begin{equation}
    s=4m_\pi^2,\qquad t=0,\qquad u=0.\label{eq:thpipi}
\end{equation}
Similar to the $\pi K$ scattering, the corresponding $S$-wave scattering lengths at the threshold are
\begin{equation}
    a_0^I
    =
    \frac{T^I(4m_\pi^2,0,0)}{32\pi},
    \qquad I=0,\ 2.\label{eq:pipil}
\end{equation}

Comparing Eq. \eqref{eq:piKA} with Eq. \eqref{eq:pipiA}, it shows that the $\pi\pi$ amplitude is subject to a stronger symmetry constraint than that of $\pi K$. This difference will be useful later when the chiral power-counting behavior of the two processes is compared.

\section{Numerical Results for $\pi K$ and $\pi\pi$ Scatterings}\label{sec:num}

The $\mathcal{O}(p^2)$ result for the $\pi K$ scattering amplitude was obtained at an early stage \cite{Griffith:1968jaz}. The first systematic analysis at the $\mathcal{O}(p^4)$ order was presented in Refs.~\cite{Bernard:1990kw,Bernard:1990kx}, and the complete result at $\mathcal{O}(p^6)$ order was given in Ref.~\cite{Bijnens:2004bu}. In order to see why the power counting is breaking, it more easily starts from the numerical results.

For example, the $I=3/2$ channel scattering length $a_0^{3/2}$ is \cite{Bijnens:2014lea}
\begin{equation}
    a_0^{3/2}M_\pi=-0.071+0.007+0.016,\label{eq:a032}
\end{equation}
where the three terms in the right-hand side represent the contributions at the $\mathcal{O}(p^2)$, $\mathcal{O}(p^4)$ and $\mathcal{O}(p^6)$ orders, respectively. Evidently, the $\mathcal{O}(p^6)$-order absolute value is significantly larger than the $\mathcal{O}(p^4)$-order one, which is inconsistent with the decreasing pattern expected from the power counting rule. In order to clarify the origin of this anomalous behavior and not make the analysis complex, a more detailed topological decomposition of the $\mathcal{O}(p^6)$ amplitude is chosen below.

All $\mathcal{O}(p^6)$ calculations can be found in \cite{Bijnens:2011fm}. Their results are separated into five parts by Feynman diagrams and analytical results. For example, the $\mathcal{O}(p^6)$ isospin-$3/2$ amplitude $T^{3/2}$ can be separated into,
\begin{equation}
    T^{3/2}=T_C+T_L+T_R+T_H+T_V,\label{eq:Ts}
\end{equation}
where $T_C$ denotes the term proportional to the $\mathcal{O}(p^6)$ LECs $C_i$, i.e., the contribution generated by tree diagrams from $\mathcal{L}_6$. $T_L$ denotes the contribution proportional to the $\mathcal{O}(p^4)$ LECs $L_i$, i.e., the one-loop contribution containing vertices from $\mathcal{L}_4$. $T_R$ denotes the reducible two-loop contribution, i.e., the contribution from the two-loop diagrams, which can be factorized into a product of two one-loop integrals. $T_H$ and $T_V$ represent the parts associated with the sunset integrals $H$ and the vertex integrals $V$, respectively, but without the $T_R$ contribution, as shown in Sec.~\ref{subsection:A}. To provide a clear and intuitive impression, and also for the sake of the following discussion, Table~\ref{tab:topology_compare} lists
the contributions from the above five parts for the $\pi K/\pi\pi$ scattering amplitudes and the scattering lengths, respectively.

\begin{table}[htbp]
    \centering
    \caption{Contributions of the different topological sectors at the $\mathcal{O}(p^6)$ order for the amplitudes and scattering lengths, with the inputs in Ref. \cite{Bijnens:2014lea}.}
    \label{tab:topology_compare}
    \begin{tabular}{cccccccc}
        \hline\hline
        Processes & Quantities & $T_C$ & $T_L$ & $T_R$ & $T_H$ & $T_V$ & $T$ \\
        \hline
        $\pi K$   & $T_{\pi K}^{3/2}$ & -0.074   & 0.588   & 1.763  & 10.815 & -11.311 & 1.782 \\
        $\pi K$   & $10a_0^{3/2}M_\pi$          & -0.00642 & 0.0514  & 0.154  & 0.945  & -0.988  & 0.156 \\
        $\pi\pi$ & $A_{\pi\pi}$      & -0.0395  & 0.000287 & 1.53   & -0.468 & -0.740  & 0.283 \\
        $\pi\pi$ & $10a_0^{2}$                 & -0.0133  & -0.00321 & 0.0636 & 0.0715 & -0.120  & -0.00141 \\
        \hline\hline
    \end{tabular}
\end{table}

Table~\ref{tab:topology_compare} shows that, for $\pi K$ scattering, $T_C$ for both amplitude $T_{\pi K}^{3/2}$ and scattering length $a_0^{3/2}$ is manifestly smaller than the others. Both $T_H$ and $T_V$ are manifestly larger than others. $T_R$ and $T_L$ are between $T_C$ and $T_H$, $T_V$. However, $T_R$ is approximately equal to $T$. It seems the other terms are almost canceled. This might be an accident. Comparing with the $\mathcal{O}(p^2)$ and $\mathcal{O}(p^4)$ values in Eq. \eqref{eq:a032}, $T_C$ satisfies the power counting rule, $T_L$ can be considered approximately satisfies, but $T_R$, $T_H$ and $T_V$ all break the power counting. Hence, it seems the origin of the power-counting breaking comes from the two-loop diagrams.

On the other hand, for $\pi\pi$ scattering length $a_0^{2}$, $T_R$, $T_H$ and $T_V$ are all manifestly smaller than those from $\pi K$ scattering. However, they are also manifestly larger than the $\mathcal{O}(p^6)$-order $a_0^{2}$, as shown in Table \ref{tab:scattering_lengths}. In other words, although the two-loop contributions for $\pi\pi$ scattering is smaller than $\pi K$ scattering, they are not small enough to keep the power counting rule, respectively. $T_C$ keeps the rule. However, the sum of all five parts satisfies the power counting rule, because of a cancel. Of course, Ref.~\cite{Bijnens:2014lea} gives a very small $\mathcal{O}(p^6)$ contribution for $a_0^{2}$. It enlarges the cancellation. Refs. \cite{Yang:2020eif,Pan:2023qja} give larger $\mathcal{O}(p^6)$ values, as shown in Table \ref{tab:scattering_lengths}. The influence from $T_C$ and $T_L$ are mitigated, but the two-loop contribution is also a little large.

Hence, in any case, the two-loop diagrams may cause the power-counting breaking. The next section will investigate the calculation and begins from the two-loop diagrams.

\section{Origin of the Chiral Power-Counting Breaking}\label{sec:origin}

This section examines the calculations from the various $\mathcal{O}(p^6)$ contributions, in order to identify the mechanism responsible for the power-counting breaking from $\pi K$ scattering. Section \ref{subsection:A} gives an example to examine why the sunset contribution $T_H$ could be large. This subsection also compares the sunset structure to the reducible two-loop diagrams, in order to compare their magnitudes. Section \ref{subsection:B} analyses the cancellation mechanism between the sunset contribution $T_H$ and the vertex contribution $T_V$ in detail, in order to examine their difference. Section \ref{subsection:C} gives a discussion on one-loop and tree diagrams. Finally, Section \ref{subsection:D} compares $\pi K$ scattering with $\pi\pi$ scattering and explains their difference.

\subsection{Origin of the Large Sunset Contribution $T_H$}\label{subsection:A}

The chiral-dimension decomposition of the scattering lengths is shown in Table~\ref{tab:topology_compare}.
The first issue to be clarified is why both the sunset contribution $T_H$ and the vertex contribution $T_V$ are larger than others, and how these contributions enter the total amplitude. Since the vertex and sunset topologies are highly similar in both analytical structure and numerical behavior, the sunset contribution is first taken here as a typical example for the following analysis.

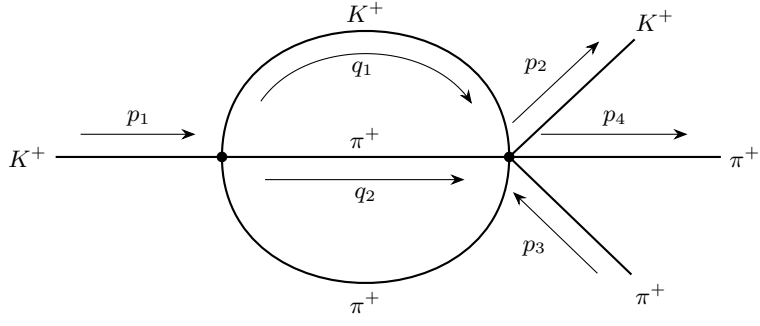
\begin{figure}[htbp]
  \centering
  \begin{tikzpicture}
    \begin{feynman}
      \vertex (v1);
      \vertex [left=2.2cm of v1] (i1) {\(K^+\)};
      \vertex [right=3.8cm of v1] (v2);

      \vertex [above right=2.2cm of v2] (f1) {\(K^+\)};
      \vertex [right=2.8cm of v2] (f2) {\(\pi^+\)};
      \vertex [below right=2.2cm of v2] (i2) {\(\pi^+\)};

      \diagram* {
        (i1) -- [plain, thick, momentum=\(p_{1}\)] (v1),
        (v1) -- [plain, thick, half left, looseness=1.5, momentum'=\(q_{1}\), edge label=\(K^+\)] (v2),
        (v1) -- [plain, thick, momentum'=\(q_{2}\), edge label=\(\pi^+\)] (v2),
        (v1) -- [plain, thick, half right, looseness=1.5, edge label'=\(\pi^+\)] (v2),
        (v2) -- [plain, thick, momentum=\(p_{2}\)] (f1),
        (v2) -- [plain, thick, momentum=\(p_{4}\)] (f2),
        (i2) -- [plain, thick, momentum=\(p_{3}\)] (v2),
      };
      \fill (v1) circle (2pt);
      \fill (v2) circle (2pt);
    \end{feynman}
  \end{tikzpicture}
  \caption{A typical sunset diagram for the $I=\frac{3}{2}$ channel. }
  \label{fig:sunset_diagram}
\end{figure}

One typical sunset diagram is shown in Fig.~\ref{fig:sunset_diagram}. Its amplitude structure is analyzed as an example, in order to clarify why the sunset integral gives a relatively large numerical contribution. The other sunset diagrams should give a similar conclusion.

The explicit expression for the scattering amplitude of Fig.~\ref{fig:sunset_diagram} is
\begin{equation}
\begin{aligned}
\mathcal{M}_{H1} = \frac{1}{432 F_0^6 M_\pi^4} & \int \frac{d^4 q_1}{(2\pi)^4} \frac{d^4 q_2}{(2\pi)^4} \frac{1}{(q_1^2 - M_\pi^2) (q_2^2 - M_\pi^2) ((-p_1 + q_1 + q_2)^2 - M_K^2)} \\
& \times \Big[ -2(p_1 \cdot q_1)(11 M_K^2 + 7 M_\pi^2 + 2s - 7t - 4u) - 2(p_1 \cdot q_2)(M_K^2 + 21 M_\pi^2 - 10s - 8t + 7u) \\
& \quad + M_\pi^2 (2M_\pi^2 - 14(q_1 \cdot q_2) - 4q_1^2 - 16q_2^2 + 2s - 7t - 10u) \Big] \\
& \times \Big[ M_\pi^2 (-2 M_K^2 + 4(q_1 \cdot q_2) - 2q_1^2 + 4q_2^2 + s + t + u) \\
& \quad + 2(p_1 \cdot q_1)(M_\pi^2 + s + t - u) + 4(p_1 \cdot q_2)(2M_\pi^2 - s - t + u) \Big].
\end{aligned}
\end{equation}
The analytical result of the numerator can be classified mainly into two types. One type contains the scalar products between the external and loop momenta, such as $p_1\!\cdot q_1$ and $p_1\!\cdot q_2$. The other type contains scalar products involving only loop momenta, such as $q_1\!\cdot q_2$, $q_1^2$, and $q_2^2$. The magnitude of these two types are discussed separately below.

An explicit example of the first type is
\begin{equation}
    \begin{aligned}
\mathcal{M}_{H1}^{(1)} = \frac{11 M_K^4}{108 F_0^6 M_\pi^2} \int \frac{d^4 q_1}{(2\pi)^4} \frac{d^4 q_2}{(2\pi)^4} \frac{p_1 \cdot q_1}{(q_1^2 - M_\pi^2)(q_2^2 - M_\pi^2)((-p_1 + q_1 + q_2)^2 - M_K^2)}.
\end{aligned}
\label{eq:M1}
\end{equation}
Using the tensor reduction scheme \cite{Amoros:1999dp}, the sunset integral can be represented by the following $H$ integrals,
\begin{align}
H\left(m_1^2,m_2^2,m_3^2;p^2\right)&=\left\langle\left\langle1\right\rangle\right\rangle,\\
H_\mu\left(m_1^2,m_2^2,m_3^2;p^2\right)&=\langle\langle q_\mu\rangle\rangle=p_\mu H_1\left(m_1^2,m_2^2,m_3^2;p^2\right),\\
H_{\mu\nu}\left(m_1^2,m_2^2,m_3^2;p^2\right)&=\langle\langle q_\mu q_\nu\rangle\rangle=p_\mu p_\nu H_{21}\left(m_1^2,m_2^2,m_3^2;p^2\right)+g_{\mu\nu}H_{22}\left(m_1^2,m_2^2,m_3^2;p^2\right),\\
\langle\langle X\rangle\rangle_H&\equiv\frac{1}{i^2}\int\frac{d^dq}{\left(2\pi\right)^d}\frac{d^dr}{\left(2\pi\right)^d}\frac{X}{\left(q^2-m_1^2\right)\left(r^2-m_2^2\right)\left(\left(q+r-p\right)^2-m_3^2\right)}.
\end{align}
In order to compare the numerical results in the same units, all $H$ integrals are dimensionless by $m_i=M_i/\mu$, where $\mu=770$ MeV. $p$, $q$ and $r$ are all dimensionless, too.
Hence, Eq.~(\ref{eq:M1}) can be reduced to
\begin{equation}
\mathcal{M}_{H1}^{(1)} = \frac{11 M_K^6}{108 F_0^6 M_\pi^2} H_1(M_\pi^2, M_\pi^2, M_K^2; p_1^2).
\label{eq:MH1}
\end{equation}

An explicit example of the second type is
\begin{align}
\mathcal{M}_{H1}^{(2)}
& = \frac{2 M_K^2}{27 F_0^6} \int \frac{d^4 q_1}{(2\pi)^4} \frac{d^4 q_2}{(2\pi)^4} \frac{q_2^2}{(q_1^2 - M_\pi^2)(q_2^2 - M_\pi^2)((-p_1 + q_1 + q_2)^2 - M_K^2)}\label{eq:MH20} \\
&= \frac{2 M_K^2}{27 F_0^6} \left[ A(M_\pi^2) A(M_K^2) + M_\pi^2 H(M_\pi^2, M_\pi^2, M_K^2; p_1^2) \right],\label{eq:MH2}
\end{align}
where $A(m^2)$ is the standard one-loop tadpole integral,
\begin{equation}
A(m^2)=\frac{1}{i}\int\frac{d^d q}{(2\pi)^d}\frac{1}{q^2-m^2}.
\end{equation}
The first term in Eq. \eqref{eq:MH2} contains a product of two $A$, which contributes the $T_R$ term in Eq. \eqref{eq:Ts}. Only the second term contributes $T_H$.

\begin{figure}[htbp]
    \centering
    \includegraphics[width=0.95\textwidth]{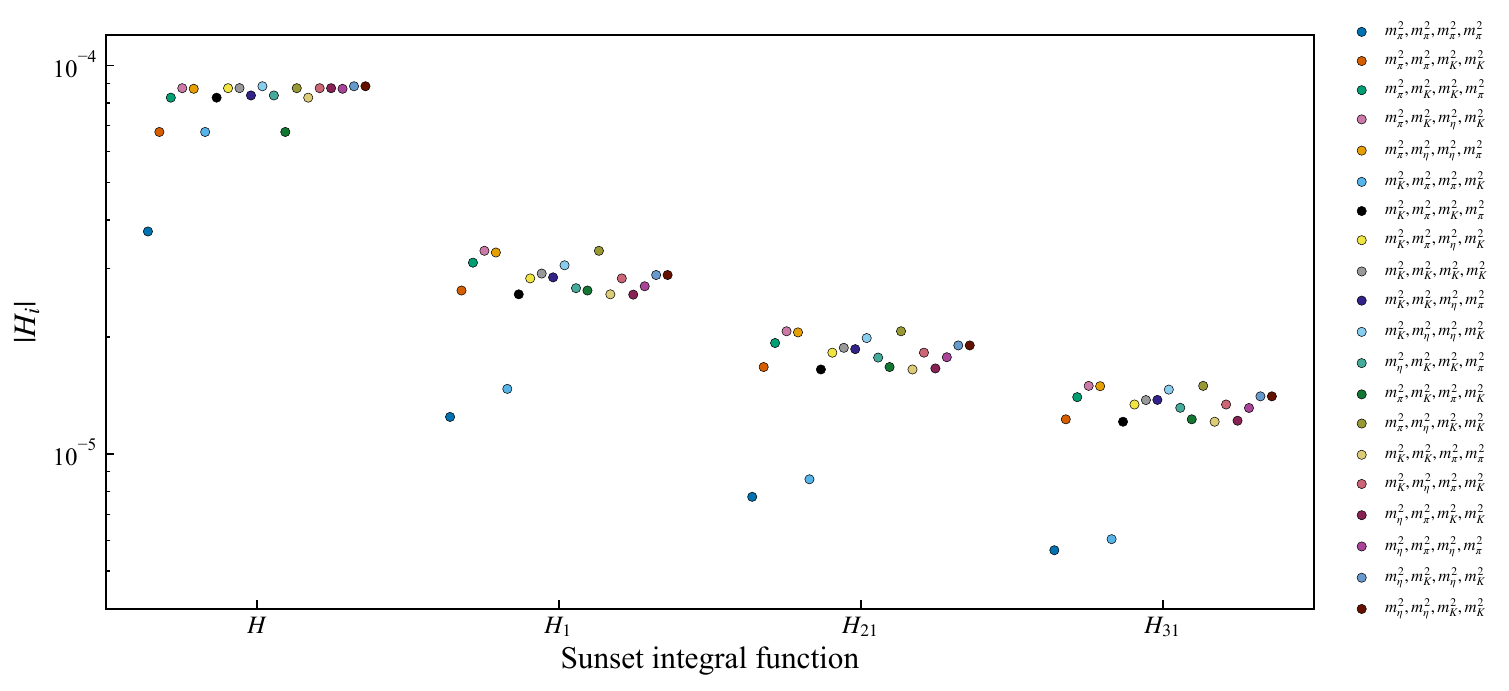}
    \caption{Distribution of the absolute values of the sunset integral functions for relevant masses parameters}
    \label{fig:sunset}
\end{figure}

The explicit numerical values of these sunset integrals are computed to high precision using the C++ package CHIRON \cite{Bijnens_2015}.
As illustrated in Fig.~\ref{fig:sunset}, the absolute values of the sunset integrals range from approximately $5.7\times10^{-6}$ to $8.8\times10^{-5}$ for the mass configurations considered. For each function, the values for the different mass configurations remain within the same order of magnitude. Nevertheless, their contributions to the amplitude also depend on the accompanying masses and kinematic coefficients.

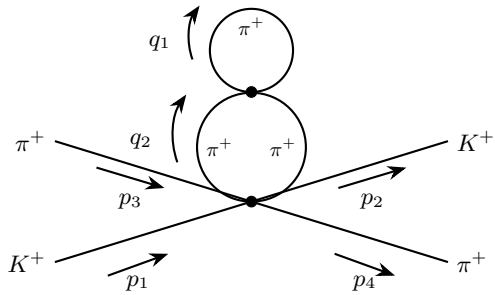
\begin{figure}[htbp]
  \centering
  \begin{tikzpicture}[scale=1.0]

    \coordinate (v0) at (0,0);
    \coordinate (v1) at (0,1.45);

    \coordinate (inK)   at (-2.6,-0.8);
    \coordinate (inPi)  at (-2.6, 0.8);
    \coordinate (outK)  at ( 2.6, 0.8);
    \coordinate (outPi) at ( 2.6,-0.8);

    \draw[thick] (inK) -- (v0);
    \draw[thick] (inPi) -- (v0);
    \draw[thick] (v0) -- (outK);
    \draw[thick] (v0) -- (outPi);

    \node[left]  at (inK)   {\(K^+\)};
    \node[left]  at (inPi)  {\(\pi^+\)};
    \node[right] at (outK)  {\(K^+\)};
    \node[right] at (outPi) {\(\pi^+\)};

    \draw[-{Stealth[length=2.5mm]}, thick] (-1.90,-0.98) -- (-1.10,-0.69);
    \node[below] at (-1.50,-0.86) {\(p_1\)};

    \draw[-{Stealth[length=2.5mm]}, thick] (-2.05,0.45) -- (-1.15,0.18);
    \node[below] at (-1.60,0.22) {\(p_3\)};

    \draw[-{Stealth[length=2.5mm]}, thick] (1.15,0.18) -- (2.05,0.45);
    \node[below] at (1.60,0.22) {\(p_2\)};

    \draw[-{Stealth[length=2.5mm]}, thick] (1.10,-0.69) -- (1.90,-0.98);
    \node[below] at (1.50,-0.86) {\(p_4\)};

    \coordinate (c1) at (0,2.00);
    \draw[thick] (c1) circle (0.55);
    \node at (0,2.32) [scale=0.85] {\(\pi^+\)};
    \draw[-{Stealth[length=2.5mm]}, thick] (-0.78,1.88) arc[start angle=200,end angle=145,radius=0.78];
    \node[left] at (-0.95,2.1) {\(q_1\)};

    \coordinate (c2) at (0,0.72);
    \draw[thick] (c2) circle (0.72);
    \node at (-0.42,0.72) [scale=0.85] {\(\pi^+\)};
    \node at ( 0.42,0.72) [scale=0.85] {\(\pi^+\)};
    \draw[-{Stealth[length=2.5mm]}, thick] (-0.98,0.52) arc[start angle=200,end angle=145,radius=0.98];
    \node[left] at (-1.2,0.8) {\(q_2\)};

    \filldraw[black] (v0) circle (2pt);
    \filldraw[black] (v1) circle (2pt);

  \end{tikzpicture}
  \caption{A typical reducible two-loop self-energy correction diagram.}
  \label{fig:reducible_diagram}
\end{figure}

For comparison with the irreducible sunset diagram, the reducible two-loop diagrams are considered, which contributes $T_R$ in Table~\ref{tab:topology_compare}, is small enough. One typical reducible two-loop diagram is shown in Fig.~\ref{fig:reducible_diagram}.

Its amplitude is
\begin{equation}
    \begin{aligned}
\mathcal{M}_{R1} = -\frac{1}{108 F_0^6 M_\pi^2} & \int \frac{d^4 q_1}{(2\pi)^4} \frac{d^4 q_2}{(2\pi)^4} \frac{(2 M_\pi^2 - 6(q_1 \cdot q_2) - q_1^2 - q_2^2)}{(q_1^2 - M_\pi^2) (q_2^2 - M_\pi^2)^2} \\
& \times \Big[ 2(p_1 \cdot q_2)(10 M_K^2 + 4 M_\pi^2 + 6s - 5t - 5u) \\
& \quad + M_\pi^2 (28 M_K^2 + 26 M_\pi^2 + 6q_2^2 - 16s - 7t - 4u) \Big].
\end{aligned}\label{eq:reducible}
\end{equation}

An explicit example of a term involving only scalar products of loop momenta, i.e. $q_1^2$, is
\begin{equation}
\begin{aligned}
\mathcal{M}_{R1}^{(1)} &= \frac{4s}{27 F_0^6} \iint \frac{d^4 q_1}{(2\pi)^4} \frac{d^4 q_2}{(2\pi)^4} \frac{q_1^2}{(q_1^2 - M_\pi^2)(q_2^2 - M_\pi^2)^2} \\
&= \frac{4s M_\pi^2}{27 F_0^6} A(M_\pi^2) B( M_\pi^2, M_\pi^2, 0),
\end{aligned}
\end{equation}
where $B(m_1^2,m_2^2,p^2)$ is the standard one-loop two-point integral, defined as
\begin{equation}
B(m_1^2,m_2^2,p^2)=\frac{1}{i}\int\frac{d^d q}{(2\pi)^d}\frac{1}{(q^2-m_1^2)\bigl((q-p)^2-m_2^2\bigr)}.
\end{equation}

An explicit example of a mixed scalar-product term containing an odd number of loop momenta, such as $(p_1\!\cdot q_2)\,q_2^2$
\begin{equation}
\mathcal{M}_{R1}^{(2)} = \frac{s}{9 F_0^6 M_\pi^2} \iint \frac{d^4 q_1}{(2\pi)^4} \frac{d^4 q_2}{(2\pi)^4} \frac{(p_1 \cdot q_2)\, q_2^2}{(q_1^2 - M_\pi^2)(q_2^2 - M_\pi^2)^2} = 0,
\end{equation}
where the vanishing result follows from the odd parity under $q_2 \to -q_2$.

A reducible two-loop diagram can essentially be factorized into the product of two one-loop integrals, namely the $A$ and $B$ functions. The asymptotic analytic behavior of the one-loop integrals and the sunset integral $H$ are approximately given by
\begin{align}
A(m^2) ={}& -\frac{m^2}{16\pi^2}\ln(m^2)+\cdots\label{eq:Aa}\\
B\left(m_1^2,m_2^2,p^2\right)={}& -\frac{1}{16\pi^2}\frac{m_1^2\ln(m_1^2)-m_2^2\ln(m_2^2)}{m_1^2-m_2^2}+\cdots,
\label{37}\\
H\Big(m_1^2,m_2^2,m_3^2;0\Big) ={}&  -\frac{1}{(16\pi^2)^2}\Big(m_1^2\left(\frac1{12}\pi^2+\frac32-\ln_1+\frac12(-\ln_2\ln_3+\ln_1\ln_4)\right)\notag\\
&+m_2^2\left(\frac1{12}\pi^2+\frac32-\ln_2+\frac12(-\ln_1\ln_3+\ln_2\ln_4)\right)\notag\\
&+m_3^2\Big(\frac1{12}\pi^2+\frac32-\ln_3+\frac12(-\ln_1\ln_2+\ln_3\ln_4)\Big)\Big)+\cdots,
\label{38}
\end{align}
where $\ln_i=\ln(m_i^2)$ $(i=1,2,3)$, In the present calculation, $m_i$ can take the values $m_\pi$, $m_K$, or $m_\eta$, depending on which meson running in the loop. The logarithm $\ln_4$ is defined as $\ln_4=\ln_1+\ln_2+\ln_3$. ``$\cdots$'' denotes $\epsilon$ terms in the dimension regularization. More detailed analytic structures of these loop integrals can be found in Refs.~\cite{Amoros:1999dp,Bijnens:2002hp}. Terms contain $\epsilon$ almost give the same order contribution of the constant terms. Hence, the following analysis only considers the right-hand constant terms in Eqs. \eqref{eq:Aa} -- \eqref{38}.

The origins of the product $A\cdot B$ and the sunset integral $H$ are different, i.e. the former mainly originates from the reducible two-loop diagrams, whereas the latter is a genuine irreducible two-loop integral. Nevertheless, their leading analytic structures share the same schematic form in Eqs. \eqref{eq:Aa} -- \eqref{38}. For example, both of them contain the two-loop suppression factor $(16\pi^2)^{-2}$, are proportional to meson-mass squared terms, and involve single and double chiral logarithms. These can be estimated by power counting rule and Feynman integral. In particular, Eqs. \eqref{eq:Aa} and \eqref{37} lead the product $A\cdot B$ to generate a term proportional to
\begin{align}
\frac{m^2}{(16\pi^2)^2}
\ln(m^2)
\frac{m_1^2\ln(m_1^2)-m_2^2\ln(m_2^2)}
{m_1^2-m_2^2},
\end{align}
which contains a product of two logarithms. This is structurally analogous to the logarithmic part of $H$ in Eq. \eqref{38}, where terms such as $\ln_i\ln_j$ and $\ln_i\ln_4$ appear multiplied by two $\ln m_i^2$ and one $m_i^2$. This similarity suggests that the corresponding numerical contributions should not give a large difference between these two types of integral.
However, the large different values in Table~\ref{tab:topology_compare} indicates that the final topological contributions are not determined by this integral structure alone.

After the tensor reduction of the sunset integral, most terms, such as Eq. \eqref{eq:MH20}, do not have symmetry to produce zero results. This is because the denominator has the sum of $q_1$ and $q_2$. However, for the reducible two-loop diagrams, such as Eq. \eqref{eq:reducible}, denominator only contain $q^2_1$ and $q^2_2$. Hence, in the reduction of the $A$- and $B$-type one-loop structures, many terms vanish because of parity or odd powers of loop momenta integrate. However, the structures of both sunset diagrams and reducible two-loop diagrams are very similar. Hence, the total number of terms between them, before integrating internal momenta, are almost the same. This causes the non-vanish terms in the reducible two-loop diagrams to be less than sunset diagrams, i.e. the absolute value of sunset diagrams are larger.

\subsection{Cancellation Mechanism Between the $H$ and $V$ Terms}\label{subsection:B}

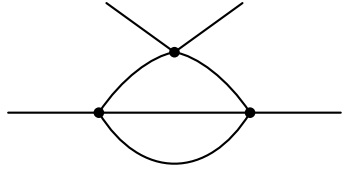
\begin{figure}[htbp]
\centering
\begin{tikzpicture}[thick, line cap=round, line join=round]

  \coordinate (L)  at (-2.2, 0);
  \coordinate (A)  at (-1.0, 0);
  \coordinate (B)  at ( 1.0, 0);
  \coordinate (R)  at ( 2.2, 0);
  \coordinate (C)  at ( 0.0, 0.8);
  \coordinate (U1) at (-0.9, 1.45);
  \coordinate (U2) at ( 0.9, 1.45);

  \draw (L) -- (A);
  \draw (A) -- (B);
  \draw (B) -- (R);

  \draw (A) .. controls (-0.7,0.45) and (-0.3,0.75) .. (C);
  \draw (C) .. controls ( 0.3,0.75) and ( 0.7,0.45) .. (B);

  \draw (A) .. controls (-0.45,-0.9) and (0.45,-0.9) .. (B);

  \draw (C) -- (U1);
  \draw (C) -- (U2);

  \fill (A) circle (2pt);
  \fill (B) circle (2pt);
  \fill (C) circle (2pt);
\end{tikzpicture}
\caption{Generic vertex Feynman diagram}
\label{fig:vertex_topology}
\end{figure}

The vertex topology is closely analogous to the sunset topology discussed above. Both of them are irreducible two-loop structures and involve coupled loop momenta. The main difference is that the vertex topology contains an additional propagator and therefore leads to a more complicated integral structure. For the later discussion, the generic form of the vertex diagram is shown in Fig.~\ref{fig:vertex_topology}.

The vertex integral $V$ has a structure closely analogous to that of the sunset integral $H$. Its standard definition is~\cite{Bijnens:2002hp}
\begin{equation}
    \begin{aligned}
        \langle\langle X\rangle\rangle_V=\frac{1}{i^2}\int\frac{d^dr}{(2\pi)^d}\frac{d^ds}{(2\pi)^d}\frac{X}{(r^2-m_1^2)((r-q)^2-m_2^2)(s^2-m_3^2)((r+s-p)^2-m_4^2)}.
    \end{aligned}\label{eq:vi}
\end{equation}
Using the method developed in Refs. \cite{Ghinculov_1995,Ghinculov_1998}, the $V$ integral can be simplified further by Feynman parameterization,
\begin{equation}
    \begin{aligned}
        \langle\langle X\rangle\rangle_V=&\int_0^1dy\frac{1}{i^2}\int\frac{d^dr}{(2\pi)^d}\frac{d^ds}{(2\pi)^d}\frac{X}{[(r-yq)^2-((1-y)m_1^2+ym_2^2-y(1-y)q^2)]^2}\\
        &\times\frac{1}{(s^2-m_3^2)((r-yq+s-(p-yq))^2-m_4^2)}.
    \end{aligned}
\end{equation}
After shifting the integration momenta, the core integral structure can be written as
\begin{equation}
    \begin{aligned}
        P_{211}^{ij}(m^2,m_3^2,m_4^2,k^2)=\frac{1}{i^2}\int\frac{d^dr}{(2\pi)^d}\frac{d^ds}{(2\pi)^d}\frac{(r\cdot k)^i(s\cdot k)^j}{(r^2-m^2)^2(s^2-m_3^2)((r+s+k)^2-m_4^2)}.
    \end{aligned}
    \label{41}
\end{equation}
Compared with the sunset integral $H$, the vertex integral $V$ contains one additional propagator in the denominator. This feature will play a crucial role in the following analysis.

For one typical case, the approximate analytic structure of Eq.~(\ref{41}) is
\begin{equation}
\begin{aligned}
P_{211}^{20}(1,2,3;k^2) \sim
\frac{1}{(16\pi^2)^2} k^2 \Bigg[
&\frac{m_1^2}{2}
\left(
\frac{1}{8}
+\frac{\pi^2}{12}
+\ln_1^2
-\frac{1}{2}\ln_1
\right)
+\frac{m_2^2}{4}
\left(
\frac{7}{8}
+\frac{\pi^2}{12}
+\ln_2^2
-\frac{3}{2}\ln_2
\right) \\
&+\frac{m_3^2}{4}
\left(
\frac{7}{8}
+\frac{\pi^2}{12}
+\ln_3^2
-\frac{3}{2}\ln_3
\right)
+\frac{1}{2}m_1^2 h_{1;123}
+\frac{1}{4}m_2^2 h_{1;213}
+\frac{1}{4}m_3^2 h_{1;312}
\Bigg]+\cdots,
\label{eq:P}
\end{aligned}
\end{equation}
where ``$\cdots$'' denotes terms contain $\epsilon$ in the dimension regularization. $h_i$ is a finite dimensionless function defined through a one-dimensional integral representation in Ref.~\cite{Ghinculov_1998}. Schematically, they have the form
\begin{equation}
    h_i=\int_0^1 dx\,\mathcal{F}_i(x),
\end{equation}
where $\mathcal{F}_i(x)$ is built from Spence functions, logarithms, and dimensionless mass ratios. Away from threshold singularities, one may estimate
\begin{equation}
    \mathcal{F}_i(x)\sim \mathcal{O}(1\text{ -- }10),
    \qquad
    h_i\sim \int_0^1 dx\,\mathcal{O}(1\text{ -- }10)
    \sim \mathcal{O}(1\text{ -- }10).
\end{equation}
Therefore, the $h_i$ terms only modify the dimensionless coefficients in the square bracket and do not change the order of magnitude of Eq.~\eqref{eq:P}.

\begin{figure}[htbp]
    \centering
    \includegraphics[width=0.95\textwidth]{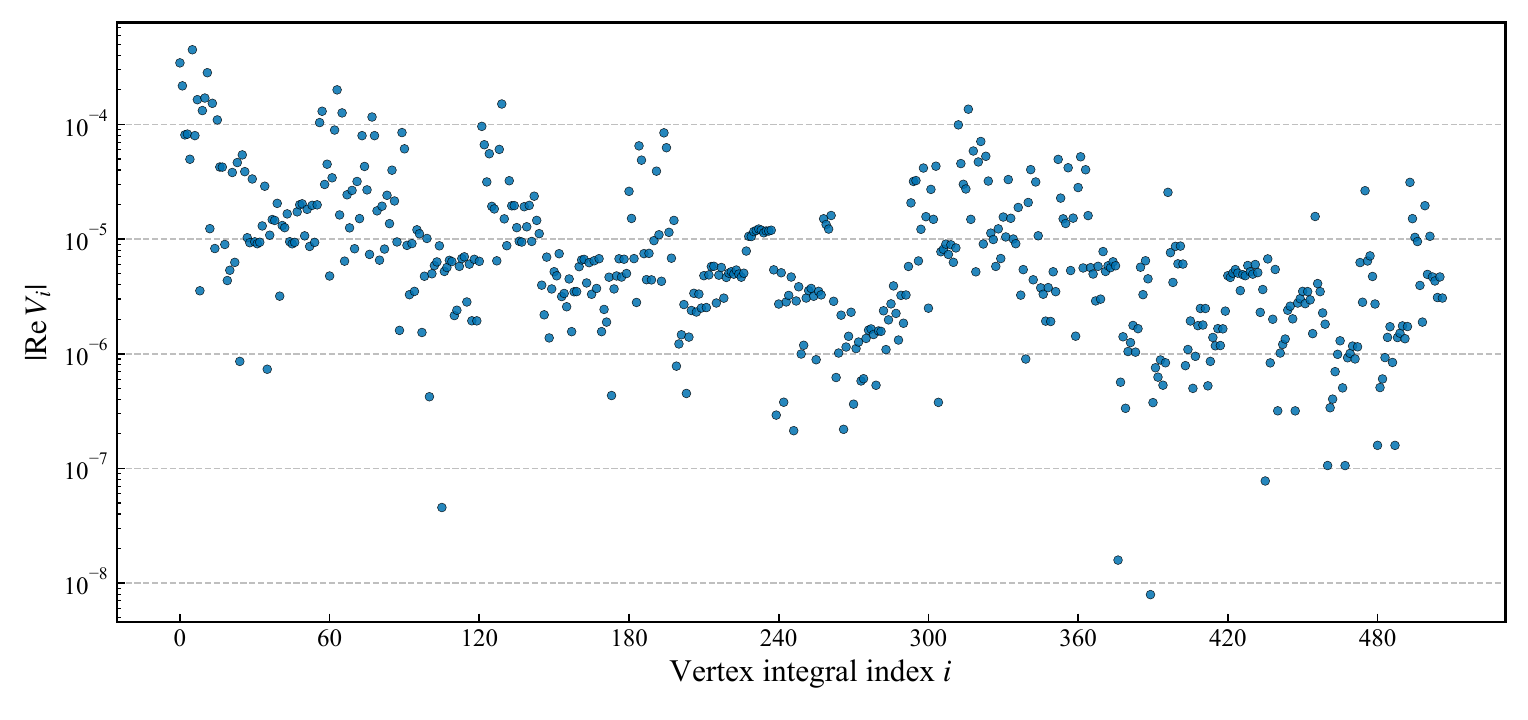}
    \caption{Numerical distribution of the vertex integrals. The horizontal axis denotes the vertex integral index $i$ and the vertical axis shows $|\mathrm{Re}\,V_i|$.}
    \label{fig:vertex}
\end{figure}

Using this estimate, Eq.~\eqref{eq:P} can be written schematically as
\begin{equation}
    P_{211}^{20}(1,2,3;k^2)
    \sim
    \frac{k^2}{(16\pi^2)^2}
    \left[
    m_1^2\mathcal{O}(1\text{ -- }10)
    +m_2^2\mathcal{O}(1\text{ -- }10)
    +m_3^2\mathcal{O}(1\text{ -- }10)
    \right].\label{eq:Ps}
\end{equation}
For a typical scale of meson mass $m$, this gives
\begin{equation}
    P_{211}^{20}
    \sim
    \frac{k^2m^2}{(16\pi^2)^2}
    \mathcal{O}(1\text{ -- }10).
\end{equation}
This should be compared with the sunset $H$ integral in Eq.~\eqref{38}, whose leading analytic structure can be estimated as
\begin{equation}
    H
    \sim
    \frac{m^2}{(16\pi^2)^2}
    \left[
    \mathcal{O}(1)
    +\mathcal{O}(\ln_i)
    +\mathcal{O}(\ln_i\ln_j)
    \right]
    \sim
    \frac{m^2}{(16\pi^2)^2}
    \mathcal{O}(1\text{ -- }10),\label{eq:Hs}
\end{equation}
where for the typical scale of meson mass $m_i$, the logarithmic terms are $\mathcal{O}(1)$ order.
Thus, at the level of magnitude estimation, the typical vertex integral contains an additional factor of the external momentum squared, i.e.~$k^2$, compared to the sunset integral.
In the low-energy region, $k^2<1$. Hence, this additional factor naturally suppresses the vertex integral.
This leads the magnitude of a vertex integral is usually smaller than that of a sunset integral.
Fig.~\ref{fig:vertex} gives the distribution of all vertex integrals by CHIRON. Most vertex integrals are smaller than corresponding sunset integrals, which confirm the above estimate.
There exist some exceptions, which come from the specific values of each item in Eqs. \eqref{eq:Ps} and \eqref{eq:Hs}. However, their values are not very large compared with Fig.~\ref{fig:sunset}. Comparing between Figs.~\ref{fig:vertex_topology} and Fig.~\ref{fig:sunset_diagram}, it shows that the vertex integrals have a higher degree of symmetry than sunset integrals, although vertex integrals are more complex.
In the actual calculation, the number of vertex integrals (507) is more than one order of magnitude larger than the number of sunset integrals (74). In other words, although the rough estimate $|T_V|$ is smaller than $|T_H|$, the actual $|T_V|$ is a little larger than $|T_H|$. Finally, the total contributions associated with $|T_H|$ and $|T_V|$ can still reach the same order of magnitude.

From the perspective of Wick rotation, the power of $-1$ in the integral coefficient is controlled directly by the number of propagators,
\begin{equation}
\int\frac{d^d\ell}{(2\pi)^d}\frac{1}{[\ell^2-\Delta]^m}=(-1)^m\frac{i}{(2\pi)^d}\int d^d\ell_E\frac{1}{[\ell_E^2+\Delta]^m}.
\end{equation}
The negative signs of the $T_H$ and $T_V$ contributions mainly from the number of internal lines in the Feynman diagrams, i.e. the number of propagators of the additional propagator in the vertex topology. Each propagator produces one $-1$, but the definition of $T_V$ has one more propagator than $T_H$, as shown in Eqs. \eqref{eq:M1} and \eqref{eq:vi}. Hence, in the majority situation, $T_V$ has one additional negative sign than $T_H$. Their positive and negative signs are different, but the absolute values are nearly equivalent, as shown in Table~\ref{tab:topology_compare}.

\subsection{One-loop and Tree Diagrams}\label{subsection:C}

For the one-loop contribution, the amplitude can be schematically written as a sum of the following term
\begin{equation}
    L_i \cdot B(m_1^2,m_2^2,p^2) \cdot \tilde{M}^L,
    \label{44}
\end{equation}
where $L_i$ is the $\mathcal{O}(p^4)$-order LECs, $B$ is a one-loop integral and $\tilde{M}^L$ is a mass coefficient, which comes from the external line.

However, for the two-loop contribution, using the sunset topology as an example, the amplitude has the general structure
\begin{equation}
     H(m_1^2,m_2^2,m_3^2;p^2) \cdot \tilde{M}^H,
\end{equation}
where $\tilde{M}^L$ is a mass coefficient as $\tilde{M}^L$.
A comparison between Eq.~\eqref{38} and Eq.~\eqref{37} shows that, comparing to the one-loop integral $B(m_1^2,m_2^2,p^2)$, the sunset integral $H(m_1^2,m_2^2,m_3^2;p^2)$ contains an additional factor of order $1/(16\pi^2)$, which is about $10^{-2}$. The sunset integral is therefore expected to be smaller than the one-loop integral by about two orders of magnitude. However, Eq. \eqref{44} has an additional $L_i\sim 10^{-3}$, which finally reduces one-loop contribution by approximately one order of magnitude than two-loop one. This is consistent with Table~\ref{tab:topology_compare}.

For the tree-level contribution, the amplitude depends linearly on the $\mathcal{O}(p^6)$-order dimensionless LECs $C_i^r$~\cite{Bijnens:2004bu}
\begin{equation}
    C_i^r \cdot \tilde{M}^C,
\end{equation}
where $\tilde{M}^C$ is the corresponding mass coefficient and $C_i^r\sim10^{-6}$~\cite{Bijnens:2014lea,Jiang:2010wa,Jiang:2015dba}.
In Table~\ref{tab:topology_compare}, $|T_C|$ is particularly small, compared to $|T_H|$. If one wants to make $|T_C|$ larger to cancel $T_R$, $T_H$ and $T_V$, in order to decrease the total $\mathcal{O}(p^6)$ contribution, one or some $C_i^r$ should be increased by at least one order of magnitude.
In actually, Refs. \cite{Pan:2023qja,Yang:2020eif,Bijnens:2014lea} obtained $C_i^r$ by a global fit method, not only considered the $\pi K$ scattering lengths. Concretely, $C_i^r$ randomly samples within a certain range (about $10^{-6}$) in a specific manner and with a certain probability. There are 27 $C_i^r$ related to $\pi K$ scattering. If one or several of these $C_i^r$ increase one order of magnitude, the sampling range should be wider than the typical order of magnitude of $C_i^r$. This leads the stochastic algorithm hard to find the $C_i^r$ values, and the other $C_i^r$ would change largely. However, the other theoretical and experimental results restrict the ranges of $C_i^r$. In addition, in the global fit, the other processes also make an impact on these $C_i^r$. These $C_i^r$ cannot be adjusted arbitrarily. It should lead to the other processes obtaining bad results. Hence, the tree diagrams provide $|T_C^{\pi\pi}|$ with a small value and keep the power counting rule.

\subsection{Compare $\pi K$ Scattering with $\pi\pi$ Scattering}\label{subsection:D}

By contrast, the $\pi\pi$ scattering length $a_0^{2}$ satisfies the power counting rule. $\pi\pi$ scattering and $\pi K$ scattering seem very similar. There is a question, why are the results so different?

From a numerical perspective, Eqs. \eqref{38} and \eqref{eq:P} indicate that small $m_\pi$ leads to small $T_H$. Because $m_\pi/m_K\sim 0.3$, $T^{\pi\pi}_H/T^{\pi K}_H\sim 0.3^2\sim 0.1$, i.e. about one order of magnitude. This is in accordance with Table \ref{tab:topology_compare}. $T_V$ gives a similar conclusion.

For one-loop diagrams, both $\pi K$ and $\pi\pi$ scattering amplitudes contain a similar structure of $L_i$ and $B(m_1^2,m_2^2,p^2)$, as Eq. \eqref{44}. The major difference comes from $\tilde{M}^L$. Dimensional analysis tells us that they are polynomials of the same degree. The maximal difference comes from the terms with the highest power of $m_K$. Because $\pi K$ scattering contains two more kaons in the external lines than $\pi\pi$ scattering, i.e.
\begin{align}
\max\frac{\tilde{M}^L_{\pi K}}{\tilde{M}^L_{\pi\pi}}\sim \bigg(\frac{M_K}{M_\pi}\bigg)^2\sim 10,
\end{align}
the one-loop contribution for $\pi K$ scattering is about one order of magnitude than that for $\pi\pi$ scattering. Because the structure of amplitudes between $\pi K$ and $\pi\pi$ scattering are similar. The final one-loop results should sum a series term as Eq. \eqref{44}. The difference between all one-loop contribution form $\pi K$ and $\pi\pi$ scattering is not change, still $|T_L^{\pi K}/T_L^{\pi\pi}|\sim 10$. It is still about one order of magnitude. This is consistent with Table~\ref{tab:topology_compare}.

From an analytical perspective, because all external lines are identical pion, $\pi\pi$ scattering have a higher degree of symmetry than $\pi K$ scattering. Hence, the number of $\pi\pi$ diagrams are smaller than the number of $\pi K$ cases. In addition, $\pi K$ scattering needs a larger symmetry factor than $\pi\pi$ scattering. All of These lead to the values of $T_R$, $T_H$ and $T_V$ being much smaller in the $\pi\pi$ scattering. Of course, the trends of their decreases may be different.

As a check, Table \ref{tab:k to pi} substitutes $M_k=M_\pi$ in the $\pi K$ scattering. Compared with Table \ref{tab:topology_compare}, although the total value is reduced, the decrement is still very small. Each value of $T_R$, $T_H$ and $T_V$ changes small. It indicates that large $M_k$ is not the major origin of the power-counting breaking. It makes only a very small contribution. In other words, although the mass of strange quark seems large, the $SU(3)$ chiral symmetry is lower than the $SU(2)$ one. It does not make a  great impact on the power counting rule, at least in the $\pi K$ scattering.
For $T_C$, $C_i^r$ is chosen the fitting values by the real $M_K$, but not the $M_\pi$ values. Because the $\mathcal{O}(p^6)$-order contribution of $a_0^{1/2}$ and $a_0^{3/2}$ is large in the fit, the change of this mass leads $T_C$ to give a large change.

\begin{table}[htbp]
    \centering
    \caption{Contributions of the different topological sectors at the $\mathcal{O}(p^6)$ order for the amplitude and scattering length of $\pi K$ scattering, but substitute $M_K=M_\pi$ by hand.}
    \label{tab:k to pi}
    \begin{tabular}{cccccccc}
        \hline\hline
        Process & Quantities & $T_C$ & $T_L$ & $T_R$ & $T_H$ & $T_V$ & $T$ \\
        \hline
        $\pi K$   & $T_{\pi K}^{3/2}$ & -8.195   & 1.637   & 1.295  & 16.251 & -9.336 & 1.652 \\
        $\pi K$   & $10a_0^{3/2}m_\pi$          & -0.716 & 0.143  & 0.113  & 1.420  & -0.816  & 0.144 \\
        \hline\hline
    \end{tabular}
\end{table}

\section{Summary and Discussion}\label{sec:con}

The convergence problem of ChPT has been investigated through the structure of $\pi K$ scattering amplitude and the corresponding scattering lengths.
A term-by-term tensor decomposition of the two-loop Feynman diagrams shows that both the sunset ($T_H$) and vertex ($T_V$) integrals are large, comparing to the final results. Both of them break the power counting rule. However, they have a similar integral structure, a simple estimate indicates they are different by a factor $k^2<1$, which is related to the external momentum. In addition, an extra negative sign comes from Wick rotation. These negative signs lead to $T_H$ and $T_V$ almost being canceled, but the remainder is not even small enough.
The reducible two-loop integrals $T_R$ are also compared with $T_H$, by estimating the integral structure. This estimate indicates $|T_R|$ is smaller than $|T_H|$, which confirms the numerical results.
The one-loop ($T_L$) and tree ($T_C$) contributions are also considered. The estimation indicates they are smaller than $|T_H|$, $|T_V|$ and $|T_R|$, and approximately keep the power counting rule. In short, any one of the three parts ($T_H$, $T_V$ and $T_R$) from the too-loop diagrams violate the power counting rule for $\pi K$ scattering lengths at the $\mathcal{O}(p^6)$ order.

In order to gain a different perspective on the origin of the power-counting breaking, a comparative analysis of $\pi\pi$ scattering is also given. Because $\pi\pi$ scattering has a larger symmetry than $\pi K$ scattering. Although the two-loop contribution seems even large, their violations are greatly mitigated. Hence, the previous papers can adjust the LECs $C_i^r$ in their typical range, in order to keep the power counting rule. Large $M_K$ mass also make a contribution to the power-counting breaking, but it is rather small.

As the review in the introduction, the power-counting breaking in the $\pi K$ scattering is not an isolated numerical accident, but more likely reflects a rather general dynamical mechanism. The above analysis indicates that even though all diagrams in the $\mathcal{O}(p^6)$ order are considered, the low-symmetry two-loop diagrams can break the power counting rule. Hence, when one considers other processes with diagrams like sunset or vertex diagrams, the power-counting breaking would appear. If the calculation is only up to a low order, the theoretical errors would be larger than the conventional power counting rule. If a more accurate result is wanted, the tedious high-order calculation is needed.

This paper merely gives a preliminary exploration of the power-counting breaking. There should exist other mechanisms to violate the power counting rule. There also exist many rooms for improvement on the conventional power counting rule. One possible way is to rearrange the many-loop diagrams. For example, for the problem of these $\pi K$ scattering lengths, some two-loop diagrams may be counted as the lower-order contribution, in order to keep the convergence of chiral expansion. In addition, some canceling components can be grouped together. Although each component may have large absolute value, their sum is small and can be counted as the higher-order contribution. For example, if $T_H+T_V$ could be canceled small enough, the sum should be counted as the higher order. In the framework of the conventional power counting rule, it can also give a rough analysis on the many-loop diagrams first, in order to estimate their contribution. If the estimation indicates that the contribution of these many-loop diagrams is too small to be calculated, one can choose to be lazy and not calculate these parts. All the above possibilities are beyond the scope of this paper. Some of them may be studied in future work.

\section{Acknowledgments}
Ma is deeply grateful to Professor Bijnens for generously providing the relevant C++ code. The authors also thank Jin Zhang and Qin-He Yang for their early exploration in this work. This project was supported by the Guangxi Science Foundation under Grants No.2025GXNSFAA069930, the Guangxi Science and Technology Innovation Platform Program (Leitai Action Plan, Grant No. Guike LT2600640026), Guangxi Key R\&D Program (Guangxi Funeng Action Plan, Grant No. Guike FN2504240040), and the ``Guangxi Highland of Innovation Talents'' Program.

\bibliography{refs}

\end{document}